\documentstyle[aps,preprint,epsf]{revtex}

\newcommand{\mathrm}{\mbox}

\begin{document}

\draft

\title{Comparison of the Effective Interaction to Various Orders in
       Different Mass Regions}

\author{M.\ Hjorth-Jensen}

\address{ECT*, European Centre for Theoretical Studies
in Nuclear Physics and Related Areas, I-38050 Trento, Italy}

\author{H.\ M\"{u}ther}

\address{Institut f\"{u}r Theoretische Physik, Universit\"{a}t T\"{u}bingen,
D-72076 T\"{u}bingen, Germany}

\author{E.\ Osnes}

\address{Department of Physics,
University of Oslo, N-0316 Oslo, Norway}

\author{A.\ Polls}

\address{Departament d'Estructura i Constituentes de la Materia,
Universitat de Barcelona, E-08028 Barcelona, Spain}

\maketitle

\clearpage

\begin{abstract}
The convergence of the perturbation expansion for the effective
interaction to be used in shell-model calculations is investigated
as function of the mass number $A$,
from $A=4$ to $A=208$.
As the mass number increases, there are more intermediate states
to sum over in each higher-order
diagram which contributes to the effective
interaction. Together with the fact that the energy denominators
in each diagram are smaller for larger mass numbers, these two
effects could largely enhance higher-order contributions to the
effective interaction, thereby deteriorating the order-by-order
convergence of the effective interaction. This effect is
counterbalanced by the short range of the nucleon-nucleon interaction,
which implies that its matrix elements are weaker for valence
single-particle states in ``large'' nuclei with large mass number
as compared to those in light nuclei. These effects are examined by
comparing various mean values of the matrix elements. It turns out that
the contributions from
higher-order terms remain fairly stable as the mass number
increases from $A=4$ to $A=208$. The implications for nuclear structure
calculations are discussed.

\end{abstract}


\clearpage

\section{Introduction}

One of the long-standing problems in nuclear many-body theory
has been the convergence of the perturbative expansion for the
effective interaction $H_{\mathrm{eff}}$ (or equally well that of
effective operators) derived from realistic nuclear forces to be used
in nuclear structure calculations.
Conventionally, the various
terms which appear in a perturbative expansion for $H_{\mathrm{eff}}$
are displayed by way of Feynman-Goldstone diagrams, examples
of such diagrams are shown in Fig.\ \ref{fig:fig1}. It is well known
that a realistic nucleon-nucleon interaction $V_{\mathrm{NN}}$ contains strong
components of short range, which renders
a perturbative expansion in terms of $V_{\mathrm{NN}}$ meaningless. To overcome
this problem, one takes into account the short-range correlations
through the solution of the Bethe-Brueckner-Goldstone equation and
considers a perturbation in terms of the nuclear reaction matrix $G$.
The wavy lines in Fig.\ \ref{fig:fig1} represent such $G$ interactions.
However, higher-order perturbative contributions in terms of the
$G$-matrix, may be large, and the convergence of the perturbative
expansion slow or not convergent at all.
Actually, Barrett and Kirson \cite{bk70}
showed that third-order contributions to the effective interaction
were substantial, and raised the question whether the perturbative
Rayleigh-Schr\"{o}dinger (RS)
expansion in terms of the nuclear $G$-matrix did converge at all.
Schucan and Weidenm\"{u}ller \cite{sw72} even pointed out that the
order-by-order expansion of the
effective interaction will ultimately diverge
when so-called intruder states are present. Typical intruder states
for nuclei like $^{18}$O  and $^{42}$Ca
are four-particle-two-hole
core-deformed states. It ought however to be
mentioned that for nuclei with more valence nucleons
in e.g.\ the oxygen mass area, such intruder state configurations
may not be important, and a two-body (or many-body)
effective interaction defined
within the $1s0d$-shell only, may represent the relevant degrees of
freedom.

Most  microscopic investigations of $H_{\mathrm{eff}}$ have been
performed for nuclei in the $1s0d$-shell, with few valence nucleons
outside a $^{16}$O core.
However, when one extends the area of investigation to nuclei
in the mass regions of calcium, tin or lead,
one has to face the problem that
for diagrams like those displayed in Fig.\ \ref{fig:fig1},
there are more intermediate states which contribute
to each diagram of the effective interaction
in e.g.\ the $1p0f$-shell than in the $1s0d$-shell.
Moreover, the energy spacing between the various major shells is also
smaller for nuclei in the $1p0f$-shell than for those around $^{16}$O.
This leads to smaller energy denominators which should
enhance third-order or higher-order contributions.
Thus,
the combined action of the above effects could
seriously deteriorate
the order-by-order convergence (if it does converge)
of the effective interaction. The only
mechanism which could quench these effects, is the fact that
the matrix elements of $G$ calculated in the $1p0f$-shell should in
general be weaker than those in the $1s0d$-shell. The single-particle
wave functions for the states around the Fermi energy exhibit larger
radii and, as the nucleon-nucleon interaction is of short range, the
matrix elements of $G$ should be weaker for the heavier nuclei. The
same arguments apply of course as well for the tin and lead regions.

It is then the scope of this work to study the convergence of the
effective interaction in terms of the mass number $A$, in order
to assess whether higher-order contributions to the two-body
effective interaction decrease or increase as $A$ increases.
To achieve this, we calculate all non-folded valence
linked diagrams through third-order in the interaction $G$, and
sum higher-order folded diagrams to infinite order for the mass
regions beyond closed-shell cores
with $A=4$, $A=16$, $A=40$, $A=100$ and $A=208$. The details
on how to obtain these effective interactions are briefly sketched
in the next section, together with our results and discussions.
Some concluding remarks are given in
section three.

\section{Computational details and results}

There are basically two main approaches in perturbation theory used
to define an effective operator and effective interaction,
each with its hierarchy of sub-approaches. One of these main
approaches is an energy-dependent
approach, known as Brillouin-Wigner
perturbation theory, while the Rayleigh-Schr\"{o}dinger (RS) perturbation
expansion stands for the energy independent approach. The latter is
the most commonly used approach in the literature \cite{ko90,lm85},
an approach which we will also employ here.
It is then common practice in perturbation theory to reduce the infinitely
many degrees of freedom of the Hilbert space to those represented
by a physically motivated subspace, the shell-model valence space.
In such truncations of the Hilbert space, the notions of a projection
operator $P$ on the model space and its complement $Q$ are
introduced. The projection operators defining the model and excluded
spaces are defined by
\begin{equation}
    P=\sum_{i=1}^{D}\left | \psi_i\right\rangle
      \left\langle \psi_i \right |,
\end{equation}
and
\begin{equation}
    Q=\sum_{i=D+1}^{\infty}\left | \psi_i\right\rangle
      \left\langle \psi_i \right |,
\end{equation}
with $D$ being the dimension of the model space, and $PQ=0$, $P^2 =P$,
$Q^2 =Q$ and $P+Q=I$. The wave functions $\vert{\psi_i}\rangle$
are eigenfunctions
of the unperturbed hamiltonian $H_0 = T+U$ (with eigenvalues
$\varepsilon_i$), where $T$ is the kinetic
energy and $U$ an appropriately chosen one-body potential, in this work
that of the
harmonic oscillator (h.o.).
The oscillator energies  $\hbar\Omega$ will be derived from
$\hbar\Omega = 45A^{-1/3} - 25A^{-2/3}$,  $A$ being the mass
number. This yields $\hbar\Omega =18.4$, $\hbar\Omega =13.9$,
$\hbar\Omega =11.0$, $\hbar\Omega =8.5$ and $\hbar\Omega =6.9$
MeV for $A=4$, $A=16$,
$A=40$, $A=100$ and $A=208$, respectively.
The full hamiltonian
is then rewritten as $H=H_0 +H_1$ with $H_1=V_{\mathrm{NN}}-U$,
$V_{\mathrm{NN}}$
being the
nucleon-nucleon (NN) interaction. Below we will replace $V_{\mathrm{NN}}$ by
the $G$-matrix, which will be used as the starting point
for our perturbative treatment.

Following the above philosophy, we choose
the model spaces which are believed, from
both experiment and theoretical calculations, to be relevant
for calculations of particle-particle effective
interactions in the mass areas from $A=4$ to $A=208$. These are
the $0p_{3/2}$ and $0p_{1/2}$ orbits for $H_{\mathrm{eff}}$ in the mass
area of $A=4$, the $0d_{5/2}$, $0d_{3/2}$ and $1s_{1/2}$ orbits
for $A=16$, the $1p_{3/2}$, $1p_{1/2}$, $0f_{7/2}$ and $0f_{5/2}$ orbits
for nuclei in the mass region of $A=40$ and the
$0h_{11/2}$, $0g_{7/2}$, $1d_{5/2}$, $1d_{3/2}$ and $2s_{1/2}$ orbits
for $A=100$. For these systems, the closed-shell cores ($^4$He, $^{16}$O,
$^{40}$Ca and $^{100}$Sn) have equal numbers of protons and neutrons,
and the model spaces are the same for both protons and neutrons.
For lead however, with $Z=82$ and $N=126$, the proton
and neutron model spaces are different, i.e.\ the orbits
$0i_{13/2}$, $0h_{9/2}$, $1f_{7/2}$, $1f_{5/2}$, $2p_{3/2}$ and $2p_{1/2}$
for the proton model space and
$0i_{11/2}$, $0j_{15/2}$, $1g_{9/2}$,
$1g_{7/2}$, $2d_{5/2}$, $2d_{3/2}$ and $3s_{1/2}$
for the neutron model space. Since the effective interaction theory we
will employ is tailored to degenerate model spaces, we will make
no attempt to derive for lead an effective proton-neutron
interaction for these two model spaces. Moreover, as discussed in
Ref.\ \cite{hko94}, a multishell effective interaction may show
strong non-hermiticities, or even divergencies if a h.o.\ basis
is used. Thus, for $A=4$ to $A=100$ we will discuss both isospin $T=0$
and $T=1$ effective interactions, whereas for lead we restrict
the attention to $T_z=-1$ and $T_z=1$, where $T_z$ is the projection
of the total isospin.

For the above model spaces, there are in total 15 matrix
elements for the effective interaction of $A=4$, 63 for $A=16$,
195 for $A=40$, 353 for $A=100$, 711 for the neutron model space
of $A=208$ and 353 for the proton model space of $A=208$.  The effective
interactions for $A=16$, $A=40$ and $A=100$ are listed in Ref.\
\cite{hko94}, and have been tested in nuclear structure calculations
and a good agreement with the experimental data obtained
for several isotopes in these mass areas. The spectra for isotopes
in the lead region will be published elsewhere \cite{heho95}.

Having defined the various model spaces, the next step in our
calculation is to obtain the nuclear reaction matrix $G$, given by
\begin{equation}
    G=V_{\mathrm{NN}}+V_{\mathrm{NN}}\frac{\tilde{Q}}{\omega - H_0}G,
\label{eq:betheg}
\end{equation}
where $\omega$ is the unperturbed energy of the interacting nucleons,
and $H_0$ is the unperturbed hamiltonian. For the bare NN interaction
we use the One-Boson-Exchange potential Bonn A defined
in Table A.1 of Ref.\ \cite{mac89}. The operator $\tilde{Q}$
is a projection operator which prevents the
interacting nucleons from scattering into states occupied by other nucleons.
Note that the exclusion operator used in the calculation
of the $G$-matrix in this work is different from the $Q$ operator
used in the evaluation of the effective interaction.
The definition of the Pauli operator for the $G$-matrix can be found
in Refs.\ \cite{hko94,kkko76}, where the so-called double-partitioned
scheme has been used. This means that low-lying two-particle states are
excluded by $\tilde Q$ from the intermediate states in the
Bethe-Goldstone Eq.\ (\ref{eq:betheg}). For the example of the
$1s0d$-shell this exclusion refers to states with two nucleons in the
$1p0f$-shell.
As a consequence, we have to include in our perturbation expansion
ladder type  diagrams, such as (2-3) in Fig.\ \ref{fig:fig1},
where the allowed intermediate states are those of the $1p0f$-shell or
corresponding ones for the other model-spaces.

The next step is to define the so-called $\hat{Q}$-box\footnote{Not to
be confused with the Pauli operators $Q$ of the effective interaction and
the $G$-matrix $\tilde Q$.} given by
\begin{equation}
   P\hat{Q}P=PH_1P+
   P\left(H_1 \frac{Q}{\omega-H_{0}}H_1+H_1
   \frac{Q}{\omega-H_{0}}H_1 \frac{Q}{\omega-H_{0}}H_1 +\dots\right)P,
   \label{eq:qbox}
\end{equation}
where we will replace $H_1$ with $G$ ($G$ replaces the free NN interaction
$V_{\mathrm{NN}}$).
The $\hat{Q}$-box is made up of non-folded diagrams which are irreducible
and valence linked. A diagram is said to be irreducible if between each pair
of vertices there is at least one hole state or a particle state outside
the model space. In a valence-linked diagram the interactions are linked
(via fermion lines) to at least one valence line. Note that a valence-linked
diagram can be either connected (consisting of a single piece) or
disconnected. In the final expansion including folded diagrams as well, the
disconnected diagrams are found to cancel out \cite{ko90}.
This corresponds to the cancellation of unlinked diagrams
in the Goldstone expansion \cite{ko90}.

We can then obtain an effective interaction
$H_{\mathrm{eff}}=H_0+V_{\mathrm{eff}}$ in terms of the $\hat{Q}$-box,
with \cite{ko90,hko94}
\begin{equation}
    V_{\mathrm{eff}}^{(n)}=\hat{Q}+{\displaystyle\sum_{m=1}^{\infty}}
    \frac{1}{m!}\frac{d^m\hat{Q}}{d\omega^m}\left\{
    V_{\mathrm{eff}}^{(n-1)}\right\}^m .
    \label{eq:fd}
\end{equation}
Observe also that the
effective interaction $V_{\mathrm{eff}}^{(n)}$
is evaluated at a given model space energy
$\omega$, as is the case for the $G$-matrix as well. For all mass
areas, we fix $\omega=-20$ MeV.
The first iteration is then given by
\begin{equation}
   V_{\mathrm{eff}}^{(0)}=\hat{Q}.
\end{equation}
In this work we define the $\hat{Q}$-box to consist of all
diagrams through third order in the $G$-matrix, as discussed in
Ref.\ \cite{hko94}.
Less than ten iterations were needed in order to obtain a converged
effective interaction for the various values of $A$. For further
details, see Ref.\ \cite{hko94}.
In the calculation of the
various diagrams, we limit the intermediate state excitations
to $2\hbar\omega$ in oscillator energy, an approximation which is
viable if one employs an NN  potential with a weak tensor force (such as the
Bonn potential used here),
as discussed by Sommermann {\em et al.} \cite{smtkf81}.
It is the aim of this study to explore the effects of the various
contributions to $V_{\mathrm{eff}}$. As it will be rather confusing to
discuss the effects for individual matrix elements (recall that
depending on the model-space there are up to few hundred matrix
elements), we define averages of matrix elements by
\begin{equation}
     \langle O \rangle_{diag}=\frac{
      {\displaystyle\sum_{j}\sum_{JT}}(2J+1)(2T+1)
      \left \langle j \right | O\left | j \right \rangle_{JT}}
      { {\displaystyle\sum_{j}\sum_{JT}}(2J+1)(2T+1) }\, ,
      \label{eq:aver1}
\end{equation}
where the summation index $j$ refers to all two-particle states of the
model-space under consideration, coupled to angular momentum $J$ and
isospin $T$\footnote{In tables \ref{tab:tab1} and \ref{tab:tab2} we
omit to divide with the number of configurations, as this
gives rather small numbers for the heavier nuclei.}.
In the averaging procedure defined in this equation we
have weighted the matrix elements by the factor $(2J+1)(2T+1)$ since
this factor accounts for the degeneracy of two-particle states with
respect to the projection quantum numbers and occurs e.g. in the
calculation of the energy if all valence states are occupied. It turned
out, however, that the main features of the results discussed below
are obtained as well, if this weighting factor is dropped. For the
operator $O$ we will consider $V^(1)$, which corresponds to the bare
$G$ matrix, $Q^{2}$, the $\hat Q$-box including terms up to second
order in $G$ without folded diagrams, and $V^{(2)}$ ($V^{(3)}$) the
effective interaction including all $\hat Q$-box diagrams up to second
(third) order plus all folded diagrams derived from these $\hat Q$-boxes.
Note, that the average defined in Eq.\ (\ref{eq:aver1}) includes only
diagonal matrix elements. In order to study if the conclusions remain
valid for all matrix elements we also define a mean value including all
matrix elements by
\begin{equation}
     \langle O \rangle=\frac{
      {\displaystyle\sum_{kl}\sum_{JT}}(2J+1)(2T+1)
      \left \langle k \right | O \left | l \right \rangle_{JT}}
      { {\displaystyle\sum_{kl}\sum_{JT}}(2J+1)(2T+1) }\, ,
      \label{eq:aver2}
\end{equation}
where the summation indices $k$ and $l$ include again all two-particle
states of the model-space considered. Beside these averages, which
include matrix elements of isospin $T=0$ and $T=1$, we will also report
on results where the averaging is restricted to one of these isospins
only.

Results for the mean values of diagonal matrix elements (see
Eq.\ (\ref{eq:aver1})) are listed in table \ref{tab:tab1}, while averages
including the non-diagonal matrix elements as well (see
Eq.\ (\ref{eq:aver2})) are presented in table \ref{tab:tab2} for the
various model-spaces considered.

Inspecting these tables one observes very clearly that the mean values
for the matrix elements are getting less attractive for the model
spaces referring to heavy nuclei. This trend can be observed independent
on the approximation used to calculate $V_{\mathrm{eff}}$. This
behavior reflects the fact that also the effective interaction,
calculated with inclusion of higher order terms, is of short range and
therefore, as we discussed already above, yield weaker matrix elements
for the valence nucleons in heavy nuclei as compared to the light
systems.

Furthermore, we observe some features which are valid independent on
the mass number and model space considered:
\begin{itemize}
\item The inclusion of second-order $\hat Q$-box diagrams in $Q^{(2)}$
yields a substantial attraction for the $T=0$ matrix elements and a
repulsion for $T=1$. This difference may be understood by the following
argument:
For the $T=1$ channel, the major
mechanism which accounts for the difference between
first and second order, is provided by the core-polarization diagram
in (2-2) of Fig.\ \ref{fig:fig1}.
Moreover, in the $T=1$ channel, the tensor force component of the
nucleon-nucleon interaction is not so important, whereas in the
$T=0$ channel the contribution from the $^3S_1$-$^3D_1$ partial
wave plays an important role in ladder-type diagrams, such as several
of the folded diagrams, or the particle-particle ladder
diagram in (2-3) of Fig.\ \ref{fig:fig1}. Typically, for many $J=1$ and $T=0$
particle-particle effective interactions, the particle-particle ladder
is of the size of or larger than the core-polarization diagram, while
for $J=0$ and $T=1$, the core-polarization diagram and the $G$-matrix
yield the largest contribution to the effective interaction.

\item The inclusion of folded diagrams yields a repulsive trend going
form $Q^{(2)}$ to $V^{(2)}$. The effect is again much larger in the
$T=0$ than in the $T=1$ matrix elements, which can as well be understood
from the importance of the particle-particle ladder diagrams in the
$T=0$ states. Comparing the results of $V^{(1)}$ and $V^{(2)}$ one
observes a repulsion for both isospins.

\item Contrary to this repulsion due to the second-order terms in the
folded-diagram expansion, the additional inclusion of terms of third
order in $G$ yields some attraction in $V^{(3)}$ as compared to
$V^{(2)}$. Except for the case of $^4$He, the effect of third-order
terms is very weak for the $T=1$ states.
This was also observed in Ref.\
\cite{hko94} in the study of the spectra of nuclei with valence particles
being only neutrons or protons. There the authors
noted that the spectra of e.g.\ $^{18}$O
or $^{42}$Ca obtained with either a second-order or
third-order effective interaction were quite similar.
For calculations of the
effective interaction for lead or tin,
this is a gratifying property since it
means that one needs only to evaluate  the $\hat{Q}$-box to second
order and sum all folded diagrams.
\end{itemize}

Finally, in order to discuss the convergence of the perturbation
expansion, we compare in table \ref{tab:tab1a} the ratios evaluated
from the mean values defined in Eq.\ (\ref{eq:aver1}). These ratios
reflect of course the same features which we already discussed above.
They emphasize, however, in a much better way that the different ratios
are rather insensitive on the mass number which is considered. This
means that one can expect the convergence of the perturbation expansion
for the residual interaction to be as good (or bad) for heavy nuclei as
for the light nuclei around $^{16}$O, which are usually studied.

For nuclear structure studies of heavy nuclei with neutron numbers
quite different from the proton number one typically considers
model-spaces, which are separate for protons and neutrons, ignoring the
residual interaction beyond the mean-field approximation. For these
cases (isospin $T=1$), the effects of terms of second order in $G$ seem to
be rather important with $V^{(2)}$ containing a correction of around 50
percent of the average of $V^{(1)}$. However, it is encouraging to note
that the inclusion of third order terms yields a correction of only 5
percent or even below.

\section{Conclusions}

We have studied the behavior of the perturbation
expansion for the effective interaction to be used in shell-model
studies of nuclei with various mass numbers. Inspecting appropriate
mean values of matrix elements, we have found that
the fact that the $G$-matrix becomes smaller in absolute value
with increasing mass numbers, counterbalances the effects that
there are more intermediate
states to sum over and that the energy denominators become smaller
in each individual diagram of the effective interaction.
Therefore, the convergence of the perturbation expansion seems to be
rather insensitive to the nuclear mass number. We observe that various
features of the folded-diagram expansion, which had been discussed for the
mass region $A\approx 16$, can also be found in heavy nuclei.
The nuclear structure calculations for heavy nuclei are mainly
sensitive to the proton-proton and neutron-neutron residual
interactions. For these $T=1$ matrix elements the
third-order and second-order averages are very close, indicating that for
this isospin channel one can approximate the effective interaction by
including all diagrams to second order plus folded diagrams to all orders.
For $T=0$, one still needs to account for third-order contributions.

The fact that third-order contributions seem to stabilize for heavier
nuclei, has also important consequences for nuclear structure
calculations in nuclei in the mass regions of e.g.\ $^{132}$Sn
and $^{208}$Pb. This means that the methods used to calculate
the effective interaction for valence nucleons, applied mainly
in the mass regions of $^{16}$O and $^{40}$Ca, can be applied to
the mass regions of $^{132}$Sn
and $^{208}$Pb, as done recently in Refs.\ \cite{heho95,ehhko95}.

\bigskip
We gratefully acknowledge the financial support of the NorFA (Nordic Academy
for Advanced Study), grant 93.40.018/00. One of us,
MHJ, thanks the Istituto Trentino di Cultura, Italy  and the Research
Council of Norway for their support.

\clearpage

\listoftables
\listoffigures

\clearpage

\begin{table}[hbtp]
\caption{The  mean values for diagonal matrix elements calculated
according to Eq.\ (7) in model-spaces with cores as indicated in the first
row assuming various approximations for the effective interaction.
Averages are listed for all isospins ($\sum_{T}$) as well as for
$T=0$  ($\sum_{T=0}$) and $T=1$ ($\sum_{T=1}$).
For lead, results for averages in the proton-proton model-space
($\sum_{pp}$) and the neutron-neutron model-space ($\sum_{nn}$) are
listed. All entries in MeV.}
\begin{center}
\begin{tabular}{lccc}
&\multicolumn{1}{c}{$\sum_T$}&
\multicolumn{1}{c}{$\sum_{T=0}$}&
\multicolumn{1}{c}{$\sum_{T=1}$}\\ \hline
&&&\\
$^4$He&  & & \\
$\langle V^{(1)} \rangle$&-3.42 &-5.98 &-2.23  \\
$\langle Q^{(2)} \rangle$& -3.38&-7.04 &-1.68  \\
$\langle V^{(2)} \rangle$& -2.87&-5.85 &-1.48  \\
$\langle V^{(3)} \rangle$&-3.42 &-7.21 &-1.65 \\
&&&\\
$^{16}$O& & & \\
$\langle V^{(1)} \rangle$&-1.39 &-2.67 &-0.88  \\
$\langle Q^{(2)} \rangle$& -1.24&-3.10 &-0.50  \\
$\langle V^{(2)} \rangle$&-0.99 &-2.39 &-0.43  \\
$\langle V^{(3)} \rangle$&-1.11 &-2.80 &-0.44 \\
&&&\\
$^{40}$Ca& & & \\
$\langle V^{(1)} \rangle$&-0.68 &-1.35 &-0.43  \\
$\langle Q^{(2)} \rangle$& -0.58&-1.57 &-0.21  \\
$\langle V^{(2)} \rangle$&-0.45 &-1.20 &-0.18  \\
$\langle V^{(3)} \rangle$&-0.53 &-1.51 &-0.17  \\
&&&\\
$^{100}$Sn& & & \\
$\langle V^{(1)} \rangle$&-0.28 &-0.57 &-0.17  \\
$\langle Q^{(2)} \rangle$& -0.23&-0.61 &-0.10  \\
$\langle V^{(2)} \rangle$&-0.19 &-0.48 &-0.08  \\
$\langle V^{(3)} \rangle$&-0.21 &-0.57 &-0.07 \\
&&&\\
$^{208}$Pb&  & \multicolumn{1}{c}{$\sum_{pp}$}&
\multicolumn{1}{c}{$\sum_{nn}$}\\
$\langle V^{(1)} \rangle$& &-0.09 &-0.06 \\
$\langle Q^{(2)} \rangle$& & -0.04 & -0.04 \\
$\langle V^{(2)} \rangle$& & -0.04 & -0.03 \\
$\langle V^{(3)} \rangle$& & -0.04 & -0.03 \\
\end{tabular}
\end{center}
\label{tab:tab1}
\end{table}

\begin{table}[hbtp]
\caption{Mean values for diagonal and non-diagonal matrix elements
calculated according to Eq.\ (8). Further details see caption of table I}
\begin{center}
\begin{tabular}{lccc}
&\multicolumn{1}{c}{$\sum_T$}&
\multicolumn{1}{c}{$\sum_{T=0}$}&
\multicolumn{1}{c}{$\sum_{T=1}$}\\ \hline
&&&\\
$^4$He& & &  \\
$\langle V^{(1)} \rangle$&-2.56&-3.06 &-2.32  \\
$\langle Q^{(2)} \rangle$& -2.59&-3.56 &-2.13  \\
$\langle V^{(2)} \rangle$& -2.21&-2.95 &-1.84  \\
$\langle V^{(3)} \rangle$&-2.61 &-3.69 &-2.09  \\
&&&\\
$^{16}$O& & & \\
$\langle V^{(1)} \rangle$&-0.72 &-1.13 &-0.56  \\
$\langle Q^{(2)} \rangle$& -0.68&-1.29 &-0.44  \\
$\langle V^{(2)} \rangle$&-0.54 &-0.99 &-0.36  \\
$\langle V^{(3)} \rangle$&-0.59 &-1.16 &-0.37 \\
&&&\\
$^{40}$Ca & & & \\
$\langle V^{(1)} \rangle$&-0.28 &-0.41 &-0.22  \\
$\langle Q^{(2)} \rangle$& -0.27&-0.48 &-0.19  \\
$\langle V^{(2)} \rangle$&-0.21 &-0.36 &-0.15  \\
$\langle V^{(3)} \rangle$&-0.23 &-0.46 &-0.14  \\
&&&\\
$^{100}$Sn& & & \\
$\langle V^{(1)} \rangle$&-0.13 &-0.30 &-0.07  \\
$\langle Q^{(2)} \rangle$& -0.12&-0.33 &-0.04  \\
$\langle V^{(2)} \rangle$&-0.09 &-0.26 &-0.04 \\
$\langle V^{(3)} \rangle$&-0.11 &-0.30 &-0.04  \\
&&&\\
$^{208}$Pb&  & \multicolumn{1}{c}{$\sum_{pp}$}&
\multicolumn{1}{c}{$\sum_{nn}$}\\
$\langle V^{(1)} \rangle$& &-0.03 &-0.02  \\
$\langle Q^{(2)} \rangle$& &-0.01 & -0.01  \\
$\langle V^{(2)} \rangle$& &-0.01 &-0.01  \\
$\langle V^{(3)} \rangle$& &-0.01 &-0.01  \\
\end{tabular}
\end{center}
\label{tab:tab2}
\end{table}

\begin{table}[hbtp]
\caption{Ratios of mean values for diagonal matrix elements calculated
according to Eq.\ (7). These ratios have been evaluated from mean
values with better precision than those listed in table I.}
\begin{center}
\begin{tabular}{lcc}
&
\multicolumn{1}{c}{$\langle V^{(2)} \rangle/\langle V^{(1)} \rangle$}&
\multicolumn{1}{c}{$\langle V^{(3)} \rangle/\langle V^{(2)} \rangle$}
\\ \hline
&&\\
$^4$He&  & \\
$\sum_T$ &0.84  &1.19 \\
$\sum_{T=0}$ &0.98  &1.23 \\
$\sum_{T=1}$ &0.66  &1.11 \\
&&\\
$^{16}$O & &\\
$\sum_T$ &0.71  &1.13 \\
$\sum_{T=0}$ &0.90  &1.17 \\
$\sum_{T=1}$ &0.49  &1.02 \\
&&\\
$^{40}$Ca  & &\\
$\sum_T$ &0.66  &1.18 \\
$\sum_{T=0}$ &0.89  &1.26 \\
$\sum_{T=1}$ &0.42  &0.94 \\
&&\\
$^{100}$Sn& & \\
$\sum_T$ &0.68   &1.11 \\
$\sum_{T=0}$ &0.85   &1.19 \\
$\sum_{T=1}$ &0.49  &0.94 \\
&&\\
$^{208}$Pb& &\\
$\sum_{pp}$ &0.44 &1.00 \\
$\sum_{nn}$ &0.50 &1.00 \\
\end{tabular}
\end{center}
\label{tab:tab1a}
\end{table}

\begin{figure}[hbtp]
   \setlength{\unitlength}{1mm}
   \begin{picture}(100,200)
%   \put(15,50){\epsfxsize=12cm \epsfbox{fig1.eps}}
   \end{picture}
   \caption{Different types of valence-linked diagrams. Diagram (2-1) is the
            $G$-matrix, diagrams (2-2)-(2-4) are
            second-order terms in $G$, while
            diagrams (2-5)-(2-8) are examples of third-order diagrams. }
   \label{fig:fig1}
\end{figure}

\end{document}